\documentclass[useAMS,usenatbib,usegraphicx]{mn2e}
\usepackage{times}

\title[Optical Monitoring of PKS~1510$-$089]{Optical Monitoring of
PKS~1510$-$089: A Binary Black Hole System? }

\author[J. Wu et al.]
 {J.~Wu,$^1$\thanks{E-mail: jhwu@bao.ac.cn} X.~Zhou,$^1$ B.~Peng,$^1$ J.~Ma,$^1$ Z.~Jiang$^1$ and J.~Chen$^1$\\
 $^1$National Astronomical Observatories, Chinese Academy of Sciences,
     20A Datun Road, Chaoyang District, Beijing 100012, China
}

\date{Received; Accepted}

\pagerange{\pageref{firstpage}--\pageref{lastpage}} \pubyear{2005}

\begin{document}

\label{firstpage}

\maketitle

\begin{abstract}
Three deep flux minima were observed with nearly the same time-scales and
intervals for the blazar PKS~1510$-$089 in the past few years. A binary
black hole system was proposed to be at the nucleus of this object, and
a new minimum was predicted to occur in 2002 March. We monitored this
source with a 60/90 cm Schmidt telescope from 2002 February to April.
In combination with the data obtained by \citet{xie04} in the
same period, we presented for the 2002 minimum a nearly symmetric light
curve, which would be required by an eclipsing model of a binary black
hole system. We also constrained the time-scale of the minimum to be 35
min, which is more consistent with the time-scales ($\sim42$ min) of
the three previous minima than the 89 min time-scale given by the same
authors. The wiggling miniarcsecond radio jet observed in this object is
taken as a further evidence for the binary black hole system. The `coupling'
of the periodicity in light curve and the helicity in radio jet is discussed 
in the framework of a binary black hole system.
\end{abstract}


\section{INTRODUCTION}

Blazars, as a subset of active galactic nuclei (AGN), are characterized
by rapid and strong variability in multi-wavebands. This behaviour
gives us much important information on their central physics. Variability
studies of blazars have been essential in understanding their inner
structures, radiation mechanism, and other physical processes.

PKS~1510$-$089 has been the target of many monitoring programs
\citep[e.g.,][]{villata97,xie02}. It is a flat-spectrum radio quasar at
redshift 0.361 and has a parsec-scale jet within just 3$\degr$ to our line
of sight according to \citet{homan02}. A pronounced UV-excess, a very flat
X-ray spectrum, and a steep gamma-ray spectrum are found in this object.
Its light curve is quite different from those of other blazars. Besides
fast and small amplitude variations and irregular outbursts, some brightness
minima were also observed \citep[see][and references there in]{xie02}.
In particular, three deep minima were reported by
\citet{dai01}, \citet{xie01,xie02} with
variations of 0.65~mag/41~min on 1999 June 14, 2.00~mag/41~min on 2000
May 29, and 0.85~mag/44~min on 2001 April 16, respectively. These minima
have nearly the same time-scales ($\sim42$ min) and intervals ($336\pm14$
d), and based on which, \citet{xie02} argued that the minimum was a
periodic phenomenon and proposed that there was a binary black hole (BBH)
system in the centre of PKS~1510$-$089. The minimum occurs when the
primary black hole (PBH) is eclipsed by the secondary black hole (SBH).
They also predicted that a new minimum would occur in 2002 March, and
they confirmed it with new observations \citep{xie04}.

However, the time-scale of the new minimum reported by \citet{xie04} is
89 min, which is more than two times of the time-scales of the three
previous minima. In fact, the authors had no observation for about 55
min before the minimum point \citep[see figs.~8 and 9 in][]{xie04},
making their recorded fading phase (73 min) much longer than the
brightening phase (16 min). This highly asymmetric light curve 
would be rejected by an eclipsing model of a binary system.

We also monitored this object from 2002 February to April. In combination
with the data in \citet{xie04}, we provided new constraints on the optical
variations of PKS~1510$-$089 in this period. Sect.~2 describes our
observation and data reduction procedures. The results are presented
in Sect.~3. Sect.~4 discusses the evidence for BBH from periodic light
curves and wiggling jets and Sect.~5 gives a summary.

\section{OBSERVATION AND DATA REDUCTION}
Our optical observation was performed on a 60/90 cm Schmidt telescope
located at the Xinglong Station of the
National Astronomical Observatories of China (NAOC). A Ford Aerospace
$2048\times2048$ CCD camera is mounted at its main focus. The CCD has
a pixel size of 15 $\mu$ and a field of view of $58'\times58'$, resulting
in a resolution of 1\farcs7 pixel$^{-1}$. The telescope is equipped with a
15 color intermediate-band photometric system, which covers a wavelength
range from 3000 to 10\,000 {\AA}. The telescope and photometric system
are mainly used to carry out the Beijing-Arizona-Taiwan-Connecticut (BATC)
survey and have shown their efficiency in detecting fast variability in
blazars \citep[e.g.,][]{peng03,wu05}.

\begin{figure}
\includegraphics[width=8.4cm]{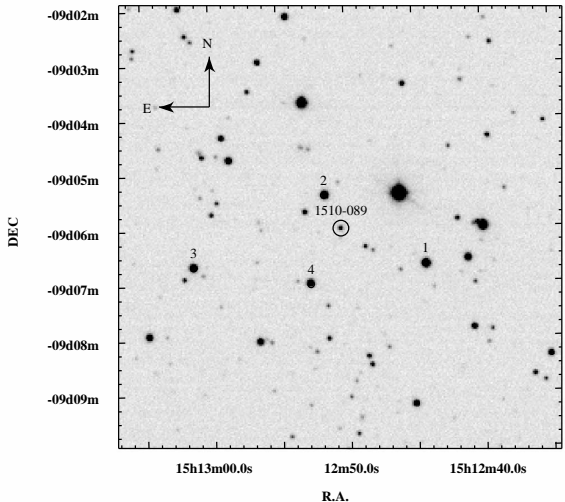}
\caption{Finding chart of PKS~1510$-$089 and four comparison stars used
by us. The image was taken with our Schmidt telescope in the BATC $i$ band
on JD~2\,452\,344 with size $8\arcmin\times8\arcmin$ (cut from a
$58\arcmin\times58\arcmin$ image).}
\label{F1}
\end{figure}

Our monitoring program covered the period from 2002 February 23 to April 10.
Three or four photometric measurements were made in each night. However,
as a result of the weather condition, only 14 night's data are useful. We
used the most sensitive $i$ filter of the BATC photometric system, which has
a central wavelength of 6711 {\AA} and a passband of 497 {\AA}, and which is 
close to the Cousins' $R$ band. An exposure time of $300\sim600$ s, depending
on the weather and seeing conditions, was able to produce a CCD image with
a good signal-to-noise ratio. The observational parameters are summarized in
Table~\ref{T1}. Fig.~\ref{F1} illustrates the finding chart of PKS~1510$-$089
and the four comparison stars used by us. The four comparison stars were also
used in other monitoring program \citep[e.g.,][]{villata97,xie02} and have
the $R$ magnitudes of $13.95\pm0.03$, $14.22\pm0.03$, $14.35\pm0.05$, and
$14.61\pm0.02$, respectively.

Two automatic procedures have been developed for the data reduction of
the BATC images. `Pipeline I' includes bias subtraction and flat-fielding
of the CCD images, and `Pipeline II' measures the instrumental magnitudes
of point sources in the BATC images. The latter is based on Stetson's standard
procedure of DAOPHOT \citep{stetson87}. The extinction coefficients
and zero points of the instrumental magnitudes were obtained by observing
the four \citet{oke83} standard stars and were used to calibrate
the instrumental magnitudes into the BATC AB magnitudes. For details of the
data reduction procedures, see \citet{zhou03}.

The light curves of PKS~1510$-$089 were given in the $R$ band in
\citet{xie04}.  For comparison, we calibrated our $i$
magnitudes into the $R$ magnitudes, which was done by using the perfect
linear relationship between the BATC $i$ magnitude and the Cousins' $R$
magnitude found by \citet{zhou03}, i.e., $R=i+0.1$. Our light curves
in the $R$ band were then obtained.

\section{Light curves and time-scale of the minimum}
\begin{figure}
\includegraphics[width=8.4cm]{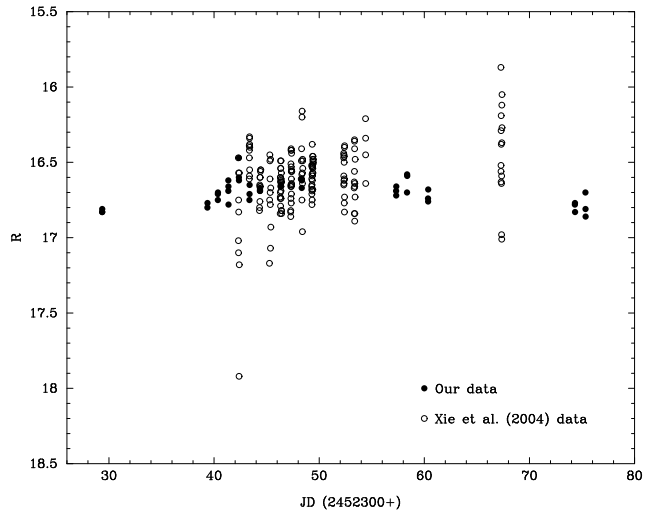}
\caption{Light curve of PKS~1510$-$089 in the $R$ band from 2002 Feb 23 to Apr
10.}
\label{F2}
\end{figure}

The observational results are given in Table~\ref{T1} with columns being the
observational date and time (UT), Julian Date, exposure time, and the $R$
magnitude and its error. The overall light curve in the $R$ band is plotted
in Fig.~\ref{F2}, in which data in \citet{xie04} are also plotted.
The two data sets are basically consistent with each other, except that
our amplitudes of variation within each night are much smaller than those of
Xie et al.'s, which is possibly due to our much shorter monitoring durations
within each night.  

\begin{figure}
\includegraphics[width=8.4cm]{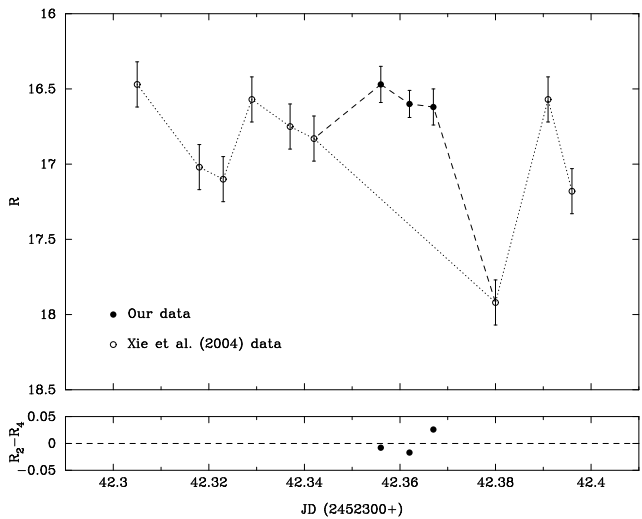}
\caption{Light curve of PKS~1510$-$089 in the $R$ band on 2002 Mar 8. The
open circles and dotted line are data in \citet{xie04} while
the filled circles and dashed line are our observations. The lower
panel gives the differential magnitudes between the 2nd
and 4th comparison stars.}
\label{F3}
\end{figure}

The predicted 2002 minimum occurred on March 8 (JD~2\,452\,342) according
to \citet{xie04}. The light curve in that night is given in Fig.~\ref{F3},
which combines our and Xie et al.'s data. The upper panel illustrates
the light curve of the target quasar while the lower panel shows the
differential magnitude (average set to 0.0) between the 2nd
and 4th comparison stars. From Fig.~\ref{F3} we can see that our
observations just filled the large gap of Xie et al.'s observations.
After a short time of fading phase beginning at JD~2\,452\,342.329,
the source underwent a brief brightening phase, as observed by us but
missed by Xie et al. Then its brightness in the $R$ band dropped
from $16.62\pm0.12$ to $17.92\pm0.15$ mag within 19 min, and
rose back to $16.57\pm0.15$ mag in 16 min. Therefore, the actual
time-scale of the minimum is 35 min, rather than 89 min reported by
\citet{xie04}. The light curve of the minimum becomes basically
symmetric, in contrary to Xie et al.'s result.

As mentioned in Sect.~1, \citet{dai01}, \citet{xie01,xie02} reported
three minima with time-scales of 41, 41, and 44 min, respectively.
\citet{xie02} argued that the minimum is a periodic phenomenon and there
is a BBH system in the centre of PKS~1510$-$089. They used the measured
time-scales to estimate the black hole masses. In this sense,
the new 35 min time-scale of the 2002 minimum,
as constrained by a combination of our and \citet{xie04} data,
is more consistent with the time-scales of the three previous minima
than the 89 min time-scale measured by \citet{xie04}. Moreover,
the basically symmetric light curve is more reasonable for an eclipsing
model.

\section{Discussion}
\subsection{BBH Model from Periodic Light Curves}
Periodicities in light curves are often attributed to the presence of
BBHs \citep[see a review by][and references therein]{komossa03}.
A prominent example is the BL Lac object OJ~287, which shows in the
optical bands quite a strict period of 11.86 yr \citep{sillan88,
sillan96,valtaoja00,pursimo00}. Most explanations of the periodicity
are based on the assumption that there is a BBH system in the nucleus
of OJ~287 \citep[e.g.,][]{sillan88,villata98,valtaoja00,liu02}. The
periodicity is linked to the orbital motion of the BBH system.

For PKS~1510$-$089, the situation is much different. Its
light curves show periodic minima rather than periodic outbursts. The
best model to explain the minima in such short time-scales is the
eclipsing model, as described by \citet{xie02}. Other possibilities
can hardly result in minima with such short time-scales. In fact,
the four minima with nearly the same time-scales and intervals and the
basically symmetric light curves would argue strongly for an eclipsing
model and a BBH system.

\begin{table*}
\centering
\begin{minipage}{10cm}
\caption{Observational log and results}
\label{T1}
\begin{tabular}{cccccc}
\hline
Observation Date & Observation Time & Julian Date & Exposure Time & $R$ & $R\rm{_{err}}$ \\
(UT) & (UT) &   & (s) & (mag) & (mag) \\
\hline
2002 02 23 & 20:26:23 & 2452329.352 & 300 & 16.83 & 0.04 \\
2002 02 23 & 20:33:58 & 2452329.357 & 300 & 16.81 & 0.04 \\
2002 02 23 & 20:41:36 & 2452329.362 & 300 & 16.83 & 0.04 \\
2002 03 05 & 20:47:40 & 2452339.366 & 600 & 16.80 & 0.05 \\
2002 03 05 & 21:00:20 & 2452339.375 & 600 & 16.77 & 0.05 \\
2002 03 06 & 20:34:59 & 2452340.358 & 300 & 16.71 & 0.05 \\
2002 03 06 & 20:42:28 & 2452340.363 & 300 & 16.75 & 0.05 \\
2002 03 06 & 20:50:17 & 2452340.368 & 300 & 16.70 & 0.05 \\
2002 03 07 & 20:28:16 & 2452341.353 & 300 & 16.69 & 0.05 \\
2002 03 07 & 20:35:48 & 2452341.358 & 300 & 16.62 & 0.05 \\
2002 03 07 & 20:43:40 & 2452341.364 & 300 & 16.66 & 0.05 \\
2002 03 07 & 20:51:08 & 2452341.369 & 300 & 16.78 & 0.06 \\
2002 03 08 & 20:33:04 & 2452342.356 & 300 & 16.47 & 0.12 \\
2002 03 08 & 20:40:47 & 2452342.362 & 300 & 16.60 & 0.09 \\
2002 03 08 & 20:48:20 & 2452342.367 & 300 & 16.62 & 0.12 \\
2002 03 09 & 20:45:13 & 2452343.365 & 300 & 16.75 & 0.03 \\
2002 03 09 & 21:01:52 & 2452343.376 & 300 & 16.71 & 0.03 \\
2002 03 09 & 21:09:24 & 2452343.382 & 300 & 16.65 & 0.03 \\
2002 03 10 & 20:35:07 & 2452344.358 & 300 & 16.69 & 0.04 \\
2002 03 10 & 20:42:37 & 2452344.363 & 300 & 16.68 & 0.04 \\
2002 03 10 & 20:50:09 & 2452344.368 & 300 & 16.67 & 0.04 \\
2002 03 10 & 20:57:50 & 2452344.373 & 300 & 16.69 & 0.04 \\
2002 03 12 & 20:07:37 & 2452346.339 & 600 & 16.66 & 0.05 \\
2002 03 12 & 20:20:17 & 2452346.347 & 600 & 16.65 & 0.04 \\
2002 03 14 & 19:56:06 & 2452348.331 & 600 & 16.61 & 0.02 \\
2002 03 14 & 20:08:45 & 2452348.339 & 600 & 16.67 & 0.02 \\
2002 03 14 & 20:21:29 & 2452348.348 & 600 & 16.62 & 0.02 \\
2002 03 23 & 19:30:26 & 2452357.313 & 600 & 16.69 & 0.03 \\
2002 03 23 & 19:43:47 & 2452357.322 & 600 & 16.72 & 0.03 \\
2002 03 23 & 19:56:18 & 2452357.331 & 600 & 16.66 & 0.03 \\
2002 03 24 & 20:32:04 & 2452358.356 & 300 & 16.70 & 0.05 \\
2002 03 24 & 20:40:39 & 2452358.362 & 300 & 16.58 & 0.05 \\
2002 03 24 & 20:48:24 & 2452358.367 & 300 & 16.59 & 0.05 \\
2002 03 26 & 20:31:41 & 2452360.355 & 600 & 16.74 & 0.06 \\
2002 03 26 & 20:44:20 & 2452360.364 & 600 & 16.76 & 0.09 \\
2002 03 26 & 20:56:59 & 2452360.373 & 600 & 16.68 & 0.07 \\
2002 04 09 & 19:38:24 & 2452374.318 & 600 & 16.78 & 0.02 \\
2002 04 09 & 19:50:52 & 2452374.327 & 600 & 16.77 & 0.02 \\
2002 04 09 & 20:03:30 & 2452374.336 & 600 & 16.83 & 0.03 \\
2002 04 10 & 19:46:14 & 2452375.324 & 600 & 16.70 & 0.05 \\
2002 04 10 & 19:58:51 & 2452375.333 & 600 & 16.81 & 0.03 \\
2002 04 10 & 20:11:27 & 2452375.341 & 600 & 16.86 & 0.03 \\
\hline
\end{tabular}
\end{minipage}
\end{table*}

\subsection{Evidence for BBH from Radio Observations}
Evidence for BBH in PKS~1510$-$089 comes also from radio observations.
PKS~1510$-$089 is a radio-loud source with high polarizations. Early radio
observations detected an arcsecond jet in this object \citep{odea88}
and classified it as a source with a well-aligned jet \citep{tingay98,cao00}.
More recently, the more sensitive VLBA and VLBI observations detected
a miniarcsecond (mas) jet in the opposite direction of its arcsecond jet
\citep{fey97,keller98,jorstad01,homan01}, and then the object shows
perhaps the most highly misaligned jet ever observed \citep{homan02}.

In order to connect the highly
misaligned mas and arcsecond jets, \citet{homan02} argued that
the jet in PKS~1510$-$089 bends at a deprojected distance of $\sim30$ kpc
from the nucleus, near the probable boundary of the host galaxy. The
bend occurs either by ram pressure from winds in the intracluster medium
or by the density gradient in the transition to the intergalactic medium.

In fact, the jet shows bends on all scales on the 1.7 and 5 GHz VLBA maps,
from mas to arcsecond. The
smaller the scale, the more significant the bend \citep[see fig.~2 in][]
{homan02}. \citet{homan02} also noted that the jet ridge-line `wiggles
in the plane of the sky after the first 10 mas'.
Similarly, the 15 GHz image of \citet{keller98} reveals a short jet
extending to the north and then bending to the northwest
within 2 mas of the core. Furthermore, multi-epoch VLBA observations have
detected several components moving away from the nucleus with different
speeds and position angles \citep[see fig.~26 in][]{jorstad01}.
This pattern of ejection naturally leads to a wiggling or precessing
inner jet.

The wiggling or precessing jet can be best explained in terms of
a BBH system. BBHs were first proposed to be in AGNs by \citet{begelman80}
and have since been frequently used to
explain the observed curved or misaligned jets \citep[e.g.,][]{roos93,
conway95}. BBH systems are formed mainly via galaxy mergers. In most
cases, the SBH is in an orbit non-coplanar with the
accretion disc of the PBH, inducing torques in the inner parts of
the disc and resulting in precession of the disc and its jet.

\begin{figure}
\begin{center}
\includegraphics[width=8.0cm]{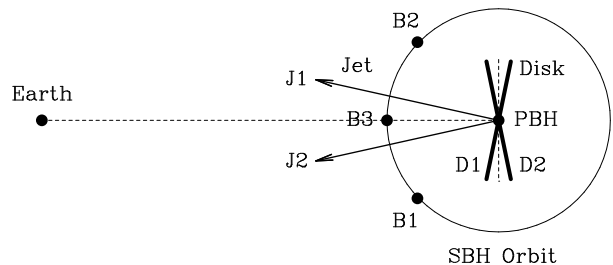}
\end{center}
\caption{Schema of the configuration of the BBH system in PKS~1510$-$089.
The Earth is in the orbital plane of the SBH. When the SBH moves along the
orbit, the inner part of the primary accretion disc will vibrate between
positions D1 and D2, and the jet will oscillate between J1 and J2, resulting
in a wiggling pattern. (The jet is not in the orbital plane and actually
follows a highly eccentric precess when the SBH moves along its orbit. See
text for details.) The angles are enlarged for clarity. The minimum occurs
when the SBH moves to position B3.}
\label{F4}
\end{figure}

Therefore, the VLBA and VLBI observations of the distorted jets in
PKS~1510$-$089 provide another evidence for the presence of the BBH in its
nucleus. However, the situation in PKS~1510$-$089 is much different. Its
light curves show periodic minima rather than periodic outbursts.
Correspondingly, it is also different in the interpretation of the periodic
minima, which is linked to the periodic eclipse of the PBH by the naked SBH
according to \citet{xie02}. In their scenario, the orbit of the SBH passes
our line of sight to the PBH. This configuration represents an extreme case
of BBH systems, for which a schema is given in Fig.~\ref{F4}. The circle
represents the orbit of the SBH around the PBH, and the Earth is in the
orbital plane. The thick lines denote two extreme positions of the inner
part of the primary accretion disc. When the SBH moves to position B1, the
inner part of the primary accretion disc tends to move to position D1 under
the gravity of the SBH, and its jet is directed to J1. When the SBH moves
to position B2, the inner part of the primary disc tends to move to position
D2 under the gravity of the SBH, and its jet is directed to J2. Therefore,
the SBH moves along the circle, makes the inner part of the primary disc
vibrating between D1 and D2 and the jet oscillating between J1 and J2.
Consequently, an apparently wiggling jet is formed. When the SBH moves to
B3, the point on our line of sight to the PBH, the PBH is eclipsed and a
flux minimum occurs. The minimum is a periodic phenomenon, and the
periodicity is caused by the orbital motion of the BBH system.

For simplicity, Fig.~\ref{F4} gives only a 2-D diagram. In the 3-D reality,
the jet will have a small angle to the orbital plane of the SBH, and the
primary accretion disc will be slightly tilted relative to the normal of
the plane. When the SBH moves along its orbit, the jet and the primary
accretion disc will follow a highly eccentric precess, with J1 and J2 being
two extreme directions of the jet. The angles in Fig.~\ref{F4} are enlarged
for clarity. The jet actually stays in roughly the same direction from the
point of view of the observer with only a small wiggling in the plane of
the sky. During the orbital motion of the BBH system, the SBH could eclipse
the PBH but it does not pass through the jet, and the jet does not cross
the line of sight when precessing.

\subsection{The `Coupling' of Periodicity and Helicity in Blazars}
It is interesting to notice that (quasi-)helical jet and periodic light
curve usually co-exist in blazars. OJ~287 itself shows evidence for a
helical jet in its 8.4 GHz VLBI map \citep{vicente96}. Other examples
presenting both helical jets and periodic light curves include
3C~120 \citep{webb90,gomez99b}, 3C~345 \citep{webb88,zensus95}, AO~0235+16
\citep{jorstad01,raiteri01}, BL Lac \citep{tateyama98,fan98}, Mrk~501
\citep{xu95,haya98}, PKS~0735+17 \citep{fan97,gomez99a}, and these objects
are most proposed to harbor a BBH system in their nuclei \citep{villata99,
caproni04a,caproni04b,ostorero04}. In other words, the periodicity in light
curve and helicity in jet usually `couple' in blazars and are interpreted
within the scenario of a BBH system. In fact, the `coupling' of the
periodicity and helicity can be easily explained in that the periodicity
in helical jets will result in periodicity in relativistic boosting in jets
and hence the periodicity in brightness.

Similarly, the periodic minima and wiggling jet observed in PKS~1510$-$089
can both be taken as evidence for a BBH system. The SBH
orbits around the PBH and exerts tidal torques on its
accretion disc, resulting in disc vibration and jet wiggling.
The eclipse of the PBH by the SBH leads to the periodic flux minima.

\section{SUMMARY}
The blazar PKS~1510$-$089 was monitored from 2002 February to April.
Based on a combination of our data and those of \citet{xie04},
we constrained a new time-scale of 35 min and a nearly symmetric light
curve for the 2002 minimum reported by Xie et al. The new results are more
consistent with the three previous minima observed by \citet{dai01},
\citet{xie01,xie02} and are more reasonable for an eclipsing model of
a BBH system suggested by \citet{xie02}. Radio observations
have revealed wiggling jets in PKS~1510$-$089 and provided another
evidence for the BBH system. The `coupling' of the periodicity in light
curve and helicity in radio jet is discussed within the framework of
BBH systems.

\section{Acknowledgments}
The authors acknowledge the anonymous referee for the insightful comments
that improved the presentation of the paper. The authors thank Professor X.
Wu for valuable discussions. This
work has been supported by the Chinese National Key Basic Research Science
Foundation (NKBRSF TG199075402) and in part by the National Natural
Science Foundation of China (NSFC), grants 10303003 and 10473012. B. Peng
acknowledges grant NKBRSF2003CB716703.

\label{lastpage}

\end{document}